\documentclass[preprint,12pt]{elsarticle}

\journal{Annals of Physics}

\RequirePackage{bm,amsfonts,lineno,hyperref,amsmath,amssymb,microtype,float}
\usepackage{epstopdf}
\usepackage{color}

\begin{document}

\title{Anisotropic compact stars: Constraining model parameters to account for physical features of tidal Love numbers}

\author{Shyam Das$^a$, Saibal Ray$^b$\footnote{$^*$Corresponding author. \\
{\it E-mail addresses:} dasshyam321@gmail.com (SD), saibal@associates.iucaa.in (SR), khlopov@apc.in2p3.fr (MK), kamalnandi1952@rediffmail.com (KKN), parida.bikram90.bkp@gmail.com (BKP).} Maxim Khlopov$^{c}$, K.K. Nandi$^d$, B.K. Parida$^e$ }

\address{$^a$Department of Physics, Malda College, Malda 732101, West Bengal, India\\
$^b$Department of Physics, Government College of Engineering and Ceramic Technology, Kolkata 700010, West Bengal, India\\
$^c$National Research Nuclear University, MEPHI (Moscow Engineering Physics Institute), \mbox{Moscow 115409, Russia}; CNRS, Astroparticule et Cosmologie, Universit\'e de Paris, F-75013 Paris, France \& Institute of Physics, Southern Federal University, 344090 Rostov on Don, Russia\\
$^d$Bashkir State Pedagogical University \& Zel'dovich International Centre for Astrophysics, Ufa 450000 (RB), Russia\\
$^e$Department of Physics, Pondicherry University, Kalapet, Puducherry 605014, India}

\maketitle

\begin{abstract}
In this paper, we develop a new class of models for a compact star with anisotropic stresses inside the matter distribution. By assuming a linear equation of state for the anisotropic matter composition of the star we solve the Einstein field equations. In our approach, for the interior solutions we use a particular form of the {\it an satz} for the metric function $g_{rr}$. The exterior solution is assumed as the Schwarzschild metric and is joined with the interior metric obtained across the boundary of the star. These matching of the metrices along with the condition of the vanishing radial pressure at the boundary lead us to determine the model parameters. The physical acceptability of the  solutions has verified by making use of the current estimated data available from the pulsar $4U~1608-52$. Thereafter, assuming anisotropy due to tidal effects we calculate the Love numbers from our model and compare the results with the observed compact stars, viz. $KS~1731-260$, $4U~1608-52$, $4U~1724-207$, $4U~1820-30$, $SAX~J1748.9-2021$ and $EXO~1745-268$. The overall situation confirms physical viability of the proposed approach, which can shed new light on the interior of the compact relativistic objects. 
\end{abstract}

\begin{keyword}
general relativity; equation of state; compact stars; hydrodynamics; anisotropy; tidal effect.
\end{keyword}

\section{Introduction}
\label{sec:1}
In General Relativity (GR), while solving the problems with compact stars, it is a general custom to assume the objects with features of spherically symmetric and isotropic nature. However, the isotropy and homogeneity of astrophysical compact stellar objects ideally may have solvable features but they need not be general physical characteristics of the stellar objects. As such the fluid pressure may have two distinct components which are responsible to provide anisotropic factor ($\Delta=p_t-p_r$) where inhomogeneity results due to the radial pressure ($p_r$) and tangential pressure ($p_t$) and hence may make the internal system of matter distribution devoid of an idealized isotropic case. This idea was first discussed by Ruderman~\cite{Ruderman1972} in his extensive review work on pulsar's structure and dynamics. Later on several scientists~\cite{Canuto1974,Bowers1974,Herrera1997} highlighted the issue in their works. However, different factors are thought to be responsible for this anisotropy, namely (i) very high density region in the core region, (ii) various condensate states (like pion condensates, meson condensates etc.), (iii) superfluid 3A, (iv) mixture of fluids of different types, (v) rotational motion, (vi) presence of the magnetic field, (vii) phase transition, (viii) relativistic particles in the compact stars etc.~\cite{Ivanov2002,SM2003,MH2003,Varela2010,Rahaman2010,Rahaman2011,Rahaman2012a,Rahaman2012b,Kalam2012,Deb2015,Shee2016,Maurya2016,Deb2017,Maurya2018,Maurya2019}.  

However, there is another factor regarding anisotropy in the compact stars that has been in speculation in the form of gravitational tidal effects. This is thought to be responsible for deformation and hence anisotropy in the fluid distribution in the stars~\cite{Hinderer2008,Doneva2012,Herrera2013,Biswas2019,Rahmansyah2020a,Roupas2020,Bhar2020,Das2020,Chatziioannou2020}. In the present work we would like to emphasize on this physical ingredient of anisotropic nature of compact stars and therefore shall be involved in calculating the Love numbers arising due to tidal effect. However, though the equilibrium configuration of the neutron star may get tidally deformed by developing a multipolar structure, in our work for the sack of simplicity, shall consider only quadrupole moment instead of multipole moment.

In this regards, we would also like to put justification of considering anisotropic fluids instead of isotropic ones. While the previous arguments put forward  by us are correct, there is a recently obtained result which somehow supersedes all previous arguments, and forces to consider pressure anisotropy whenever relativistic fluids are involved~\cite{Herrera2020}. According to this work it is a obvious fact that physical processes of the kind expected in stellar evolution will always tend to produce pressure anisotropy, even if the system is initially assumed to be isotropic. This argument reinforces further the assumption of pressure anisotropy. The important point to stress here is that any equilibrium configuration is the final stage of a dynamic regime and there is no reason to think that the acquired anisotropy during this dynamic process, would disappear in the final equilibrium state. Therefore the resulting configuration, even if  initially  had isotropic pressure, should in principle exhibit pressure anisotropy.

It is argued by Pretel~\cite{Pretel2020} that the equation of state (EOS) plays a fundamental role in determining the internal structure of such stars and, consequently, in imposing stability limits. As far as the LIGO-Virgo constraints on the EOS for nuclear matter as a result of observation of the event GW170817 is concerned it is more crucial~\cite{Pretel2020,Chatziioannou2020}. According to Fattoyev et al.~\cite{Fattoyev2018} GW170817 has provided stringent constraints on the EOS of neutron rich matter at a few times
nuclear densities from the determination of the tidal deformability of a $M =1.4~M_{\odot}$ neutron star~\cite{Bauswein2017,Abbott2018,Fattoyev2018,Annala2018,Most2018,Tews2018,Malik2018,Radice2018,Tews2019,Capano2019}. Therefore, a realistic EOS is always in demand for exploring the effects it can have on the physical characteristics of a desirable stable anisotropic compact stars. In the present investigation we are employing a linear EOS of the form $p_r = \alpha \rho +\beta$ where $\alpha$ and $\beta$ are constants so that there are ample opportunity to tune it with the observational situation for various astrophysical system of dense objects. 

Apart from the above general discussion on EOS, we would like to add here that a detailed discussions are available in refs.~\cite{Yagi2017,Rahmansyah2020b} in connection to the effects of the tidal forces and their impact on isotropic and anisotropic EOS which may be consulted by the interested authors. 

The paper is organized as follows. In Section 2 we present the associated Einstein field equations describing a spherically symmetric static anisotropic configuration. By assuming a particular form for the $g_{rr}$ metric potential and a linear EOS, we have solved the system in Section 3. The matching conditions required for the smooth connection of the interior space–time to the vacuum Schwarzschild exterior are provided in Section 4. In Section 5 we provide the bound on the model parameters required for the physical analysis of our solution. The physical viability of our model is studied in Section 6 along with the tidal deformation and Love numbers in Section 7. We conclude with some relevant discussions in Section 8.

\section{Einstein field equations} \label{sec2}
We write the line element describing the interior space-time of a spherically symmetric star in standard coordinates $x^0 = t$,  $x^1=r$,  $x^2 = \theta$,  $x^3 = \phi$ as
\begin{equation}
ds^2 = -e^{\nu(r)}(r)dt^2 + e^{\lambda(r)}dr^2 + r^2(d\theta^2 + \sin^2{\theta}d\phi^2),\label{metric}
\end{equation}
where, $e^{\nu(r)}$ and  $e^{\lambda(r)}$, are the gravitational potentials yet to be determined. 

We assume that the matter distribution of the stellar interior is anisotropic in nature and described by an energy-momentum tensor of the form 
\begin{equation}
 T_{\xi \chi} = ( \rho + p_t) u_{\xi} u_{\chi} + p_t g_{\xi \chi} - ( p_t - p_r) \eta_{\xi} \eta_{\chi},\label{eq2}
\end{equation}
where $\rho$ represents the energy-density, $p_r$ and $p_t$, respectively denote the fluid pressures along the radial and transverse directions, $u^{\xi}$ is the $4$-velocity of the fluid and $\eta^{\xi}$ is a unit space-like $4$-vector along the radial direction so that $u^{\xi} u_{\xi}= -1$, $\eta^{\xi} \eta_{\xi} = 1$ and $u^{\xi} \eta_{\xi} = 0$. Also, the above Eq. (2), for the spherically static metric with the anisotropic pressure, matches exactly with the expression of the energy-momentum-stress tensor as provided in the paper \cite{Doneva2012}. In the present work we are considering that the fluid anisotropy means that the radial pressure $p_r$ differs from the transverse pressure $p_t$.

The Einstein field equations for the line element (\ref{metric}) are obtained as (in the geometrized system of units having $ G =c=1$)
\begin{eqnarray}\label{g3}
8\pi \rho &=& \frac{\left(1 - e^{-\lambda}\right)}{r^2} + \frac{\lambda^{\prime}e^{-\lambda}}{r}, \label{g3a} \\ \nonumber \\
8\pi p_r &=&  \frac{\nu^{\prime}e^{-\lambda}}{r} - \frac{\left(1 - e^{-\lambda}\right)}{r^2}, \label{g3b}  \\  \nonumber \\
8\pi  p_t &=& \frac{e^{-\lambda}}{4}\left(2\nu^{\prime\prime} + {\nu^{\prime}}^2  - \nu^{\prime}\lambda^{\prime} + \frac{2\nu^\prime}{r} - \frac{2\lambda^\prime}{r}\right), \label{g3c}
\end{eqnarray}
where the symbol primes $(')$ represent differentiation with respect to the radial coordinate $r$.

Making use of Eqs.~(\ref{g3b}) and (\ref{g3c}), we define the anisotropic parameter of the stellar system as
\begin{align}
8 \pi  \Delta(r) &= (p_t-p_r) \nonumber \\
&= \frac{e^{-\lambda}}{4} \left[2\nu^{\prime\prime} + {\nu^{\prime}}^2  - \nu^{\prime}\lambda^{\prime} - \frac{2}{r}(\nu'+\lambda') + \frac{4}{r^2}(e^{\lambda}-1)\right].\label{eq6} 
\end{align}

Thus we have a system of four equations~(\ref{g3a})--(\ref{eq6}) with $6$ independent variables, namely $e^{\lambda}$, $e^{\nu}$, $\rho$, $p_r$, $p_t$ and $\Delta$. We need to specify two of them to solve the system. In this  model we solve the system by assuming a particular metric anasatz $g_{rr}$ and the interior matter distribution to follow a linear equation of state.

\section{Generating a physically viable model}\label{sec3}
To develop a physically reasonable model of the stellar configuration, we assume that the metric potential $g_{rr}$ is given by
\begin{equation}
e^{\lambda(r)} = 1+a r^2+b r^4,\label{etlambda}
\end{equation}
where $a$ and $b$ are the constant parameters having the units $km^{-2}$ and $km^{-4}$ respectively which are to be determined from the matching conditions. This metric potential was earlier proposed by Tolman~\cite{Tolman}  and used latter on by Biswas et al. \cite{Biswas2020} to model realistic compact stellar object. The characteristic features of this ansatz are that it is free from the central singularity and monotonic increasing function of $r$.

Interestingly, we are using the same metric component to describe relativistic anisotropic stellar objects with a prescribed linear equation of state of the form~\cite{Maurya2019a,Maurya2019b,Singh2020a,Maurya2020a,Singh2020b,Maurya2020b,Maurya2020c,Jasim2021}
\begin{equation}
p_r = \alpha \rho +\beta,  \label{eos}
\end{equation}
where $\alpha$ and $\beta$  are constants.  

Substituting assumption (\ref{etlambda}) in Eq.~(\ref{g3a}) and using Eqs.~(\ref{eos}) and~(\ref{g3b}), we have 
eventually expressions for $\nu$, $p_r$ and $p_t$, and $\rho$ respectively the metric potential, radial and tangential pressures and matter-energy density which are confined in {\bf Appendix A}.

The parameter $\beta$ can be expressed as $\beta = -\alpha \rho_R $, where $R$ is the radius of the star and $\rho_R$ is the surface density given by
\begin{equation}
\rho_R=\frac{R^2 \left(\text{a}^2+5 \text{b}\right)+2 \text{a} \text{b} R^4+3 \text{a}+\text{b}^2 R^6}{8 \pi \left(1+\text{a} R^2+\text{b} R^4\right)^2}.\label{rhoR}
\end{equation}

This ensures that the radial pressure $p_r(r=R) = 0$ whereas the central density $\rho(r=0)$ can be obtained from Eq.~(\ref{den}) of {\bf Appendix A} as
\begin{equation}
\rho_c = 3 \text{a}/8 \pi,\label{cden}
\end{equation}
which shows that we must have $ \text{a} > 0 $. The anisotropy also vanishes at the centre, i.e., $ \Delta (r=0) = 0 $.

The mass contained within a sphere of radius $r$ is defined as
\begin{equation}
\label{massfn} m(r)= 4\pi \int\limits_0^r\omega^2
\rho(\omega)d\omega,
\end{equation}
which on integration yields
\begin{equation}
m(r)= \frac{r^3(a+b r^2)}{2 (1+a r^2+b r^4)}. \label{mass}
\end{equation}

In the above expression for mass, one can note that $m(r=0) = 0$.

\section{Matching Conditions}\label{sec4}
We need to match the interior solution to the Schwarzschild exterior
\begin{equation}
 ds^2 = - \left(1 - \frac{2M}{r}\right) dt^2 + \left(1 - \frac{2M}{r}\right)^{-1}dr^2 + r^2 d\Omega^2,
\label{mm}
\end{equation}
across the boundary $R$ where $M = m(R)$ is the total mass and $d\Omega^2 = d\theta^2 + \sin^2 \theta~d\phi^2$.

The interior  metric should match smoothly to the exterior Schwarzschild spacetime metric  across the boundary of the star $r=R$. Therefore, the  continuity of the metric functions across the boundary yields
\begin{eqnarray}
e^{\nu(r=R)} &=& \left(1-\frac{2 M}{R}\right),\label{bc1}\\
e^{\lambda(r=R)} &=& \left(1-\frac{2 M}{R}\right)^{-1}.\label{bc2}
\end{eqnarray}
The radial pressure should drop to zero at a finite value of the radial parameter $r$, which is defined as the radius of the star. Utilizing the condition $p_r(r=R)=0$, we have
\begin{equation}
\frac{\alpha \left(R^2 \left(\text{a}^2+5 \text{b}\right)+2 \text{a} \text{b} R^4+3 \text{a}+\text{b}^2 R^6\right)}{8 \pi \left(\text{a} R^2+\text{b} R^4+1\right)^2}+\beta=0. \label{pr0}
\end{equation}

Fulfillment of continuity for both, the metric matching and vanishing of the radial pressure at the boundary, which is known as the junction condition has been utilized to determine the constants which can be provided below:
\begin{equation}
a=\frac{b R^5-2 M \left(b R^4+1\right)}{R^2 (2 M-R)},\label{capital_C}
\end{equation}
\begin{align}
 b &=\frac{1}{4 \alpha}\left(\frac{32 \alpha M^2-42 \alpha M R+3 (5 \alpha+1) R^2}{R^4 (R-2 M)^2}\right.\nonumber\\
&\left.-\sqrt{\frac{1}{R^8 (R-2 M)^4} \xi_{1}} \right) , \label{x}
\end{align}

\begin{align}
\beta &=\dfrac{1}{16 \pi R^4} \left[\alpha \left(-16 M^2+30 M R-15 R^2\right)+\right.\nonumber\\
&\left. R^2 \left(-3 +R^2 (-2 M + R)^2 \sqrt{\frac{1}{R^8 (R-2 M)^4} \xi_{1}}\right)\right],
\end{align}
where
\begin{align*} 
\xi_{1} =& 256 \alpha^2 M^4-960 \alpha^2 M^3 R+12 \alpha (139 \alpha+32) M^2 R^2\\
&-12 \alpha (89 \alpha+29) M R^3+9 (5 \alpha+1)^2 R^4\\
&-96 \alpha R^2 (R-2 M)^2 \left[\log (A)+\alpha \log \left(\frac{R}{R-2 M}\right)\right.\nonumber\\
&\left.-\log \left(1-\frac{2 M}{R}\right)\right].
\end{align*}

\section{Bounds on the model parameters}\label{sec5}
For a physically acceptable stellar model, it is reasonable to assume that the following conditions should be satisfied~\cite{Delgaty}:
(i) $\rho > 0$, $p_r > 0$, $p_t > 0$; (ii) $\rho' < 0$, $p_r' < 0$, $p'_t < 0$; (iii) $ 0 \leq \frac{dp_r}{d\rho} \leq 1$; $ 0 \leq \frac{dp_t}{d\rho} \leq 1$ and (iv) $\rho+p_r+2p_t > 0$. In addition, it is expected that the solution should be regular and well-behaved at all interior points of the stellar configuration. 

Based on the above requirements, bounds on the model parameters are obtained in this Section.

\begin{enumerate}
\item \textbf{Regularity Conditions:}

\begin{enumerate}
  \item The metric potentials $ e^{\lambda(r)} > 0$,~$e^{\nu(r)} > 0$~for~$0 \leq r \leq R.$\\
  
For appropriate choice of the model parameters, the above requirements are fulfilled in our model. The gravitational potentials in this model satisfy, $e^{\nu(0)}=A$ = constant, $e^{\lambda(0)}=1$, i.e., finite at the center ($r=0$) of the stellar configuration. Also one can easily check that $(e^{\nu(r)})'_{r=0}=(e^{\lambda(r)})'_{r=0}=0$. These imply that the metric is regular at the center and well behaved throughout the stellar interior which will be shown graphically.\\

 \item $ \rho (r) \geq 0, ~~ p_r (r) \geq 0, ~~ p_t (r) \geq 0 $ for $ 0 \leq r \leq R $.\\

From Eq.~(\ref{den}), we note that density remains positive if $a > 0$. Also, Eq.~(\ref{radpres}) shows that, since $dp_r/d\rho (r=0)=\alpha$ is the sound speed which must be between $0$ and $1$, so $0<\alpha <1$. From Eq. (\ref{tangpres}), we have
\begin{eqnarray}
p_t(r = 0) &= \dfrac{1}{32 \pi}\left[a \left(-4 \alpha^2+6 \alpha-2\right)+a (\alpha+1)^2\right.\nonumber\\
&\left.+(\alpha+1) (3 a \alpha+a)\right]+\beta.
\label{tangpresatcentre}
\end{eqnarray}

We note that the tangential pressure remains positive at the centre $r=0$. Fulfillment of the requirements throughout the star can be shown by graphical representation.\\

\item $p_r (r = R) = 0$. \\

From Eq.~(\ref{radpres}), we note that the radial pressure vanishes at the boundary $R$ if we set $\beta =- \alpha \rho_R $, where $\rho_R $ is the surface density. In this context we would like to give emphasis on the point that if we choose $\rho(r=R)=0$ then Eq. (8) becomes $p_r(r=R) = \beta = 0$ (because at boundary the radial pressure is zero). Also, as $\alpha$ and $\beta$ are constant parameters for a particular star, so they should remain same always. This suggests that we should not make $\rho(r=R)=0$ at the boundary, however this may be considered as a special case.
\end{enumerate}

\item \textbf{Causality Condition:}
The causality condition demands that $ 0 \leq \frac{dp_r}{d\rho} \leq 1$ and $ 0 \leq \frac{dp_t}{d\rho} \leq 1$ at all interior points of the star. Let us now represent the related expressions as follows:\\
\begin{align}
\frac{dp_r}{d\rho}&=\alpha, \\
\frac{dp_t}{d\rho}&=-\frac{\left(1+\text{a} r^2+\text{b} r^4\right)^3}{4 \left(B_1+B_2 +B_3\right)} \times \nonumber\\
	& \left[-\frac{12 (\alpha-3) \alpha r^2 \left(\text{a}^2-4 \text{b}\right) \left(\text{a}+2 \text{b} r^2\right)}{\left(1+\text{a}
   r^2+\text{b} r^4 \right)^4}\right.+\frac{4 \alpha(\alpha-3) \left(\text{a}^2-4 \text{b}\right)}{\left(1+\text{a} r^2+\text{b} r^4\right)^3} \nonumber\\
	&-\frac{\left(\text{a}+2 \text{b} r^2\right) \left(B_4+B_5r^2\right)}{\left(1+\text{a} r^2+\text{b}
   r^4\right)^2}+ \frac{4 \left(\text{a}+2 \text{b} r^2\right) \left(B_6+B_7 r^2\right)}{\left(1+\text{a} r^2+\text{b} r^4\right)^3}\nonumber\\
	& +2 \beta r^2 (\text{a} \beta+2 \text{b} (\alpha+1))+\frac{\text{b} (\alpha+1) (7 \alpha+3)-2 \text{a} (2 \alpha+1) \beta}{1+\text{a} r^2+\text{b} r^4}\nonumber\\
	& +\frac{4 \text{b}
   (\alpha (2 \alpha-5)-1)}{\left(1+\text{a} r^2+\text{b} r^4\right)^2}+\beta (2 \text{a} (\alpha+1)+\beta)+\nonumber\\
	&\left. \text{b} (\alpha+1)^2+3 \text{b} \beta^2 r^4\right],
\end{align}
where $B_1=\text{a}^3 r^2+\text{a}^2 \left(3 \text{b} r^4+5\right)$, $B_2= \text{a} \text{b} r^2 \left(3 \text{b} r^4+13\right)$, $B_3= +\text{b} \left(\text{b}^2 r^8+12 \text{b} r^4-5\right)$, $B_4=3 \text{a} \alpha^2+4 \text{a} \alpha+\text{a}-8 \alpha \beta-4 \beta$, $B_5= (\text{b} (\alpha+1) (7 \alpha+3)-2 \text{a} (2 \alpha+1) \beta)$, $B_6= \text{a} (2 \alpha^2-3 \alpha+1)$ and $B_7=2 \text{b} \left(-2 \alpha^2+5 \alpha+1\right)$.\\

At the centre $r=0$, $\frac{dp_t}{d\rho}>0$, i.e.
\begin{equation}
\frac{dp_t}{d\rho} =-\frac{\text{a}^2 (\alpha-3) (9 \alpha-1)+2 \text{a} (3 \alpha+2) \beta+40 \text{b} \alpha+\beta^2}{20 \left(\text{a}^2-\text{b}\right)}>0.
\end{equation}

Also according to Zeldovich's condition~\cite{Zeldovich1,Zeldovich2}, we must have $p_r/\rho\leq 1$ at the center. Therefore, $\frac{(3 a \alpha + \beta)}{3 a}  \leq 1$.\\

\item \textbf{Energy Condition:}
For an anisotropic fluid sphere for being physically acceptable matter composition, all the energy conditions, namely Weak Energy Condition (WEC), Null Energy Condition (NEC), Strong Energy Condition (SEC) and Dominant Energy Condition (DEC) are satisfied if and only if the following inequalities hold simultaneously in every point inside the fluid sphere:\\

NEC : $\rho + p_r \geq  0$,~$\rho + p_t \geq  0$, \\
WEC : $p_r + \rho >  0,~\rho > 0$, \\
SEC : $\rho + p_r  \geq 0,~\rho + p_r + 2 p_t \geq 0$, \\
DEC : $\rho > |p_r|,~\rho > |p_t|$.\\

We have from SEC
\begin{align}
\rho + p_r + 2 p_t (r = 0) &=\frac{3 (3 a \alpha+a+8 \pi  \beta)}{8 \pi }>0,\nonumber\\
\Rightarrow \beta & > - \dfrac{a (1 + 3 \alpha)}{8 \pi}.
\label{strngenergcondatctr}
\end{align}

As suggested by Bondi \cite{Bondi1999} and Tello-Ortiz et al. \cite{TO2020}, the Trace of the Energy Tensor (TEC) (i.e. $\rho-p_r- 2 p_t$) should be positive throughout the interior of the star. We have checked the positivity of the energy conditions TEC by plotting it graphically in Fig. 4.\\

\item \textbf{Monotonic decrease of density and pressures:}
A realistic stellar model should have the following properties:
 $ \frac{d\rho}{dr} \leq 0,~\frac{dp_r}{dr} \leq 0,~\frac{dp_t}{dr} \leq 0$ for $ 0 \leq r \leq R $. Now, we have
\begin{eqnarray}
\frac{d\rho}{dr}=-\frac{2 r(C_1+C_2+C_3)}{8 \pi \left(1+\text{a} r^2+\text{b} r^4\right)^3}, \label{drhodr}
\end{eqnarray}

\begin{eqnarray}
\frac{dp_r}{dr}=-\frac{2 \alpha r (C_1+C_2+ C_3)}{8 \pi \left(1+\text{a} r^2+\text{b} r^4\right)^3}, \label{dprdr}
\end{eqnarray}

\begin{align}
\frac{dp_t}{dr}&=\frac{1}{16 \pi} r \left[-\frac{C_4}{\left(1+\text{a} r^2+\text{b} r^4\right)^4}+\frac{C_5}{\left(1+\text{a} r^2+\text{b} r^4\right)^3}\right.\nonumber\\
& +\beta (2 \text{a} (\alpha+1)+\beta)-\frac{C_6  \left(C_7 +C_8 \right)}{\left(1+\text{a} r^2+\text{b} r^4\right)^2}+\nonumber\\
& \frac{4 C_6 (C_9+C_{10})}{\left(1+\text{a} r^2+\text{b} r^4\right)^3}+2 \beta r^2 (\text{a} \beta+2\text{b} (\alpha+1))+ \nonumber\\
	& \frac{\text{b} (\alpha+1) (7 \alpha+3)-2 \text{a} (2 \alpha+1) \beta}{1+\text{a} r^2+\text{b} r^4}\nonumber\\
&\left.  +\frac{4 \text{b} (\alpha (2 \alpha-5)-1)}{\left(1+\text{a} r^2+\text{b} r^4\right)^2}+\text{b} (\alpha+1)^2+3 \text{b} \beta^2 r^4\right], \label{dptdr}
\end{align}
\end{enumerate}
where $C_1=\text{a}^3 r^2+\text{a}^2 (3 \text{b} r^4+5)$, $C_2=\text{a} \text{b} r^2 \left(3 \text{b} r^4+13\right)$, $C_3= \text{b} \left(\text{b}^2 r^8+12 \text{b} r^4-5\right)$, $C_4= 12\left(\text{a}+2 \text{b} r^2\right) (\alpha-3) \alpha r^2 \left(\text{a}^2-4 \text{b}\right)$, $C_5= \left(\text{a}+2 \text{b} r^2\right)$, $C_6= 4 (\alpha-3) \alpha \left(\text{a}^2-4 \text{b}\right)$, $C_7=r^2 (\text{b} (\alpha+1) (7 \alpha+3)-2 \text{a} (2 \alpha+1) \beta)$, $C_8=3 \text{a} \alpha^2+4 \text{a} \alpha+\text{a}-8 \alpha \beta-4 \beta$, $C_9= \text{a} \left(2 \alpha^2-3 \alpha+1\right)$, $C_{10}= 2 \text{b} \left(-2 \alpha^2+5 \alpha+1\right) r^2$.\\

With the choices of the model parameters within their proper bound it can be  shown that both the density and radial pressure decrease radially outward. 

\begin{figure}
     \includegraphics[width=6.5cm]{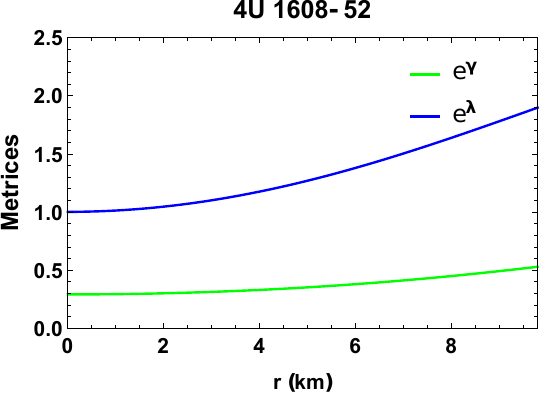}
     \includegraphics[width=6.5cm]{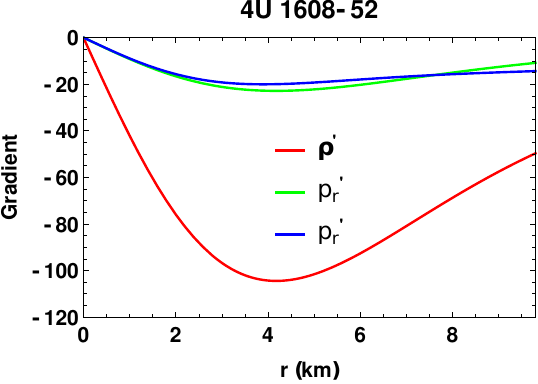}
    \caption{Variations of the metric potentials $e^\nu$ and $e^\lambda$ (left panel) whereas density and pressure gradients (right panel) are plotted against the radial coordinate $r$ inside the star.}
    \label{fig:metrices}
\end{figure}

 \begin{figure}
    \includegraphics[width=6.5cm]{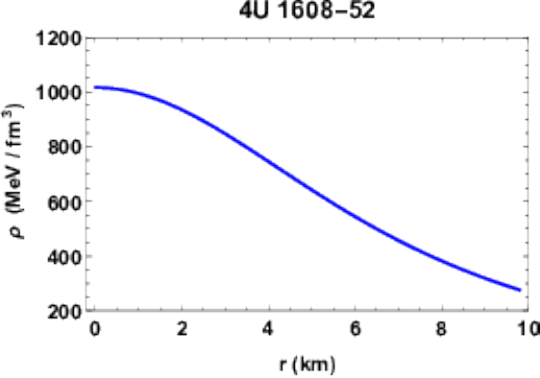}
     \includegraphics[width=6.5cm]{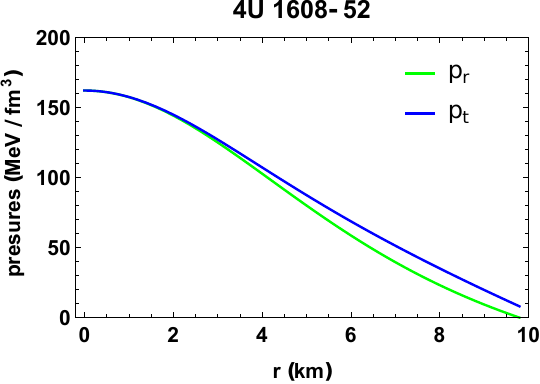}
    \caption{Fall-off behaviour of the energy density (left panel) and pressure (right panel).}
    \label{fig:Density_figure}
\end{figure}

\begin{figure}
    \includegraphics[width=6.5cm]{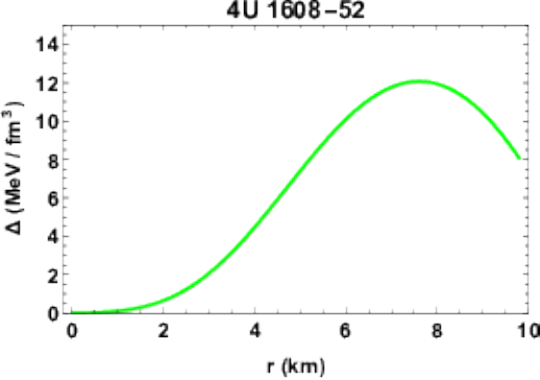}
     \includegraphics[width=6.5cm]{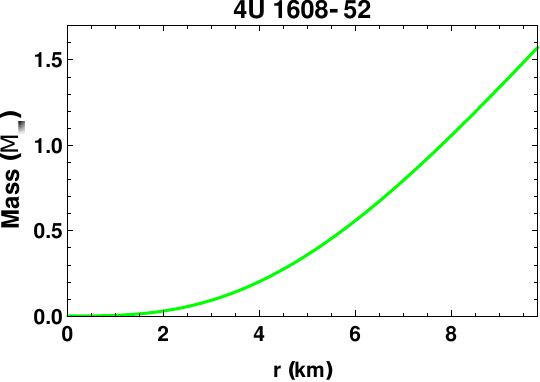}
    \caption{Radial variation of anisotropy (left panel) and stellar mass (right panel).}
    \label{fig:Ani_figure}
\end{figure}

\begin{figure}
    \includegraphics[width=6.5cm]{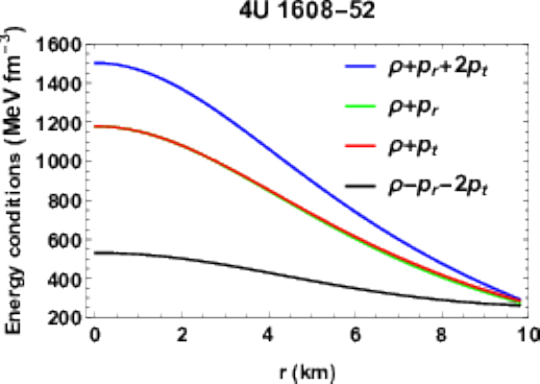}
      \includegraphics[width=6.5cm]{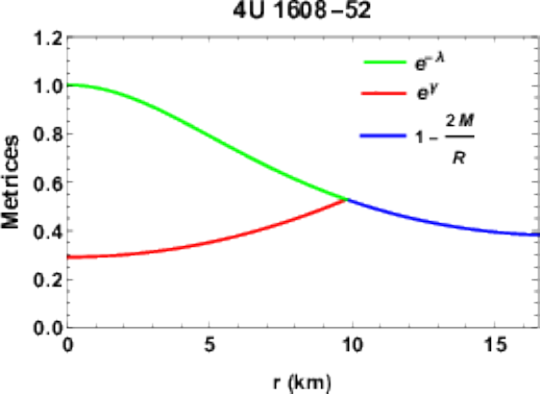}
    \caption{Verification of the energy condition (left panel) and matching of the metrices at the boundary (right panel).}
    \label{fig:Strongenergycondition_figure}
\end{figure}


\section{Stability Analysis of the model}\label{sec6} 

\subsection{TOV equation}
A star remains in static equilibrium under the forces, namely the gravitational force ($F_g$), hydrostatics force ($F_h$) and anisotropic force ($F_a$). This condition is formulated mathematically as TOV equation provided by Tolman~\cite{Tolman} and Oppenheimer and Volkoff~\cite{OV} which is described by the conservation equation as
\begin{equation}\label{tov1}
\nabla^{\mu}T_{\mu\nu}=0.
\end{equation}

Now using the expression given in (\ref{eq2}) into (\ref{tov1}) one can obtain the following equation:
\begin{equation}\label{tov3}
-\frac{\nu'}{2}(\rho+p_r)+\frac{2}{r}(p_t-p_r)=\frac{dp_r}{dr}.
\end{equation}

The Eq.~\eqref{tov3} can be written as
\begin{equation}
F_g+F_a+F_h=0,
\end{equation}

where the expressions for $F_g= -\frac{\nu'}{2}(\rho+p_r)$,~$F_a= \frac{2}{r}(p_t-p_r)$ and $F_h=-\frac{dp_r}{dr}$ are shown in {\bf Appendix B}.

The three different forces are plotted in Fig.~\ref{fig:Stabilityf_figure} (left panel) for the compact stars $4U~1608-52$. The figure shows that hydrostatics and anisotropic forces are positive and is balanced by the gravitational force which is negative to keep the system in static equilibrium.

\subsection{Adiabatic index}
The adiabatic index which is defined as
\begin{eqnarray}
\Gamma = {\rho(r)+p(r) \over p(r)}{dp(r) \over d\rho(r)},
\end{eqnarray}
is related to the stability of a relativistic anisotropic stellar configuration. 

\begin{figure}
 \includegraphics[width=6cm]{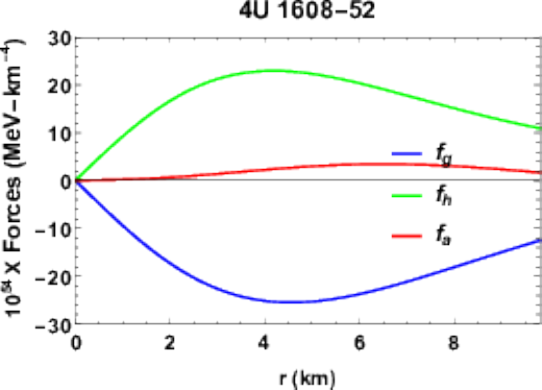}
 \includegraphics[width=6cm]{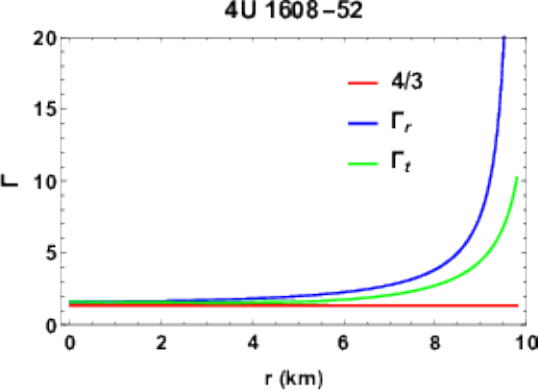}
    \caption{Variation of the forces (left panel) and adiabatic index (right panel).}
    \label{fig:Stabilityf_figure} 
\end{figure}

In fact, as it is well known, the condition for the stability of a Newtonian isotropic sphere is: $\Gamma > 4/3$~\cite{Bondi1964,Heintzmann1975,Deb2019}. This condition changes for a relativistic isotropic  sphere~\cite{Bondi1964}, and more so for an anisotropic general relativistic sphere~\cite{Herrera1979,Chan1993}. A rigorous study of dynamical stability based on the behaviour of the adiabatic index has been carried out for anisotropic fluids by Chan et al.~\cite{Chan1993}. Thus keeping all these in mind, we note that for our solution the adiabatic index $\Gamma$ takes the value more than $4/3$ throughout the interior of the compact star (as evident from the right panel of Fig.~\ref{fig:Stabilityf_figure}) and hence provides a stable configuration.

\subsection{Causality Condition}
\begin{figure}
    \includegraphics[width=6.5cm]{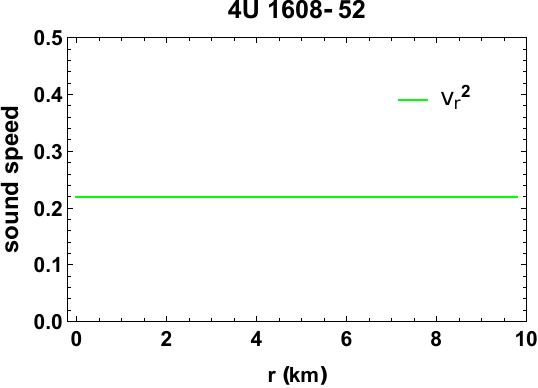}
    \includegraphics[width=6.5cm]{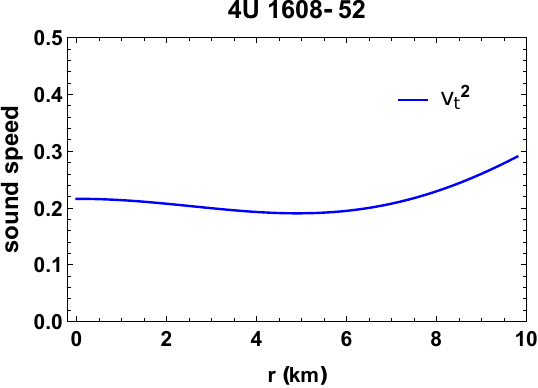}
    \caption{Verification of the radial and transverse sound speeds.}
    \label{fig:sound speed_figure}
\end{figure}

We also know that for a physically acceptable model, the velocity of the sound (both radial and transverse) should be less than the speed of the light, i.e., both $\frac{dp_r}{d\rho},~\frac{dp_t}{d\rho}<1$ which is known as the causality condition. The causality condition is shown to satisfy in Fig.~\ref{fig:sound speed_figure}.

\subsection{Harrison-Zeldovich-Novikov's stability condition}
Depending on the mass and central density of the star, Harrison et al.~\cite{Harrison} and later on Zeldovich and Novikov~\cite{ZN} proposed the stability condition for the model of any compact star. From their investigation they suggested that for stable configuration $\frac{\partial M}{\partial \rho_c}>0$, where $M$ and $\rho_c$ denote the mass and central density of the compact star.

For the present model
\begin{equation}
\frac{\partial M}{\partial \rho_c}= \frac{3 R^3}{\left(3 b R^4+3\right)^2}.
\end{equation}

\begin{figure}
\centering
    \includegraphics[width=6cm]{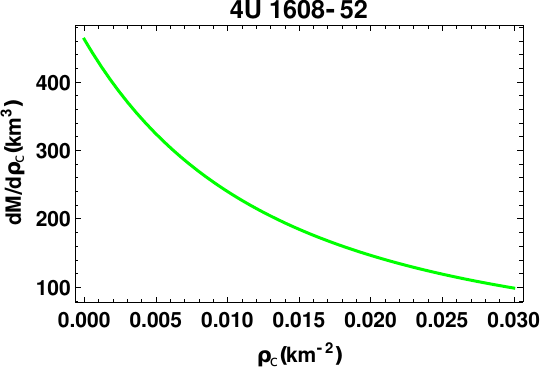}
    \caption{Variation of $dM/d\rho_c$ with respect to the central density $\rho_c$.}
    \label{fig:dmdrho_figure}
\end{figure}

Above expression of $\frac{\partial M}{\partial \rho_c}$ is always positive and hence the stability condition is well satisfied. The variation of the  $\frac{\partial M}{\partial \rho_c}$ with respect to the central density is depicted in Fig.~\ref{fig:dmdrho_figure}.

\section{Anisotropy due to tidal deformation: Evaluation of the Love number}\label{sec7} 
When a static spherically symmetric neutron star (NS) immersed in an external tidal field $\mathcal{E}_{ij}$, the equilibrium configuration of the neutron star gets tidally deformed by developing a multipolar structure. However, in our calculation for the sack of simplicity, we take only quadrupole moment $\mathcal{Q}_{ij}$ instead of multipole moment. It is because of the fact that the quadrupole moment ($l=2$) dominates over the multiple momoent if the two binary stars are sufficiently far away from each other. 

The relation between  $\mathcal{Q}_{ij}$ and  $\mathcal{E}_{ij}$ in the linear order can be written as~\cite{Hinderer2008}
\begin{eqnarray}
 \mathcal{Q}_{ij} = - \Lambda \mathcal{E}_{ij}.\label{love1}
 \end{eqnarray}
 
Here $\Lambda$ represents  the tidal deformability of the neutron star and it is related to the dimensionless tidal Love number $k_{2}$ as~\cite{Hinderer2008}
	\begin{eqnarray}
	k_{2} = \dfrac{3}{2}\Lambda \, R^{-5}.\label{love2}
\end{eqnarray}
 	
In reference to the {\bf Appendix C}, since Eq. \eqref{t20} is a first order differential equation, we need one initial condition to solve it. Setting $k_2 = 0$ as $\mathcal{C}=0$, it is clear from Eq.  \eqref{t17} that at $r = 0$, we get $y(r=0)=2$. Using this initial condition along with Eqs. \eqref{t15} and  \eqref{t16}, one can explicitly evaluate the solution of Eq. \eqref{t20}. In Figs. \ref{fig:t1}--\ref{fig:t3} the tidal Love number is plotted against $\alpha$ for different compact objects. It is evident from the plots that with increasing the value of $\alpha$ the tidal Love number decreases monotonically.

\begin{table*}[!htp]
\setlength\tabcolsep{5pt} 
\centering 
\caption{Numerical values of the physical parameters in connection to different compact stars. Here the observed masses $M$ and radii $R$ are taken from the Ref.~\cite{Ozel2016,Roupas2020}.}
\begin{tabular}{cccccccccccccccc}
\hline 
			Compact Stars & $\alpha $ & $\beta$ & A & a  & b  & M & R & $k_2$ \\ 
			 &  & ($10^{-4}$) & ($10^{-2}$) & ($km^{-2}$)  & ($km ^{-4}$)  & ($M_{\odot}$) & (km) & ($10^{-3}$)\\
			[0.5ex] 
			\hline
 			4U~1608-52 & 0.22 & -20.0  & 28.96 & 11.165 & -19.1011 & $1.57_{-0.29}^{+0.30}$ & $9.8_{-1.8}^{+1.8}$ & 38.826\\ 
 				
 			4U~1724-207 & 0.22 & -8.86173 & 28.966723 &8.7385 &-23.5788 & $1.81_{-0.37}^{+0.25}$  & $12.2_{-1.4}^{+1.4}$ & 25.9192\\
 			
 			4U~1820-30 & 0.22 & -7.24563 & 33.0 & 10.0994 & -40.2031 & $1.46_{-0.21}^{+0.21}$ & $11.1_{-1.8}^{+1.8}$ & 17.2012\\
 			
 				SAX~J1748.9-2021 & 0.22 &-12.0365 & 28.966723 & 8.64899 &   -18.3833 & $1.81_{-0.37}^{+0.25}$ & $11.7_{-1.7}^{+1.7}$ & 31.5394\\
 			
 				EXO~1745-268  & 0.22 & -16.057 &28.966723 & 10.300& -22.333& $ 1.65_{-0.31}^{+0.21}$ & $10.5_{-1.6}^{+1.6}$ & 34.4159\\
 				
 				KS~1731-260 &0.22 &   -19.593&  28.966723&  10.552 & -15.070 & $ 1.61_{-0.37}^{+0.35}$ & $10_{-2.2}^{+2.2}$ & 40.2044\\ 
\hline \label{tab1}
\end{tabular}
\end{table*}

 \begin{figure}
 \centering
\includegraphics[width=8cm]{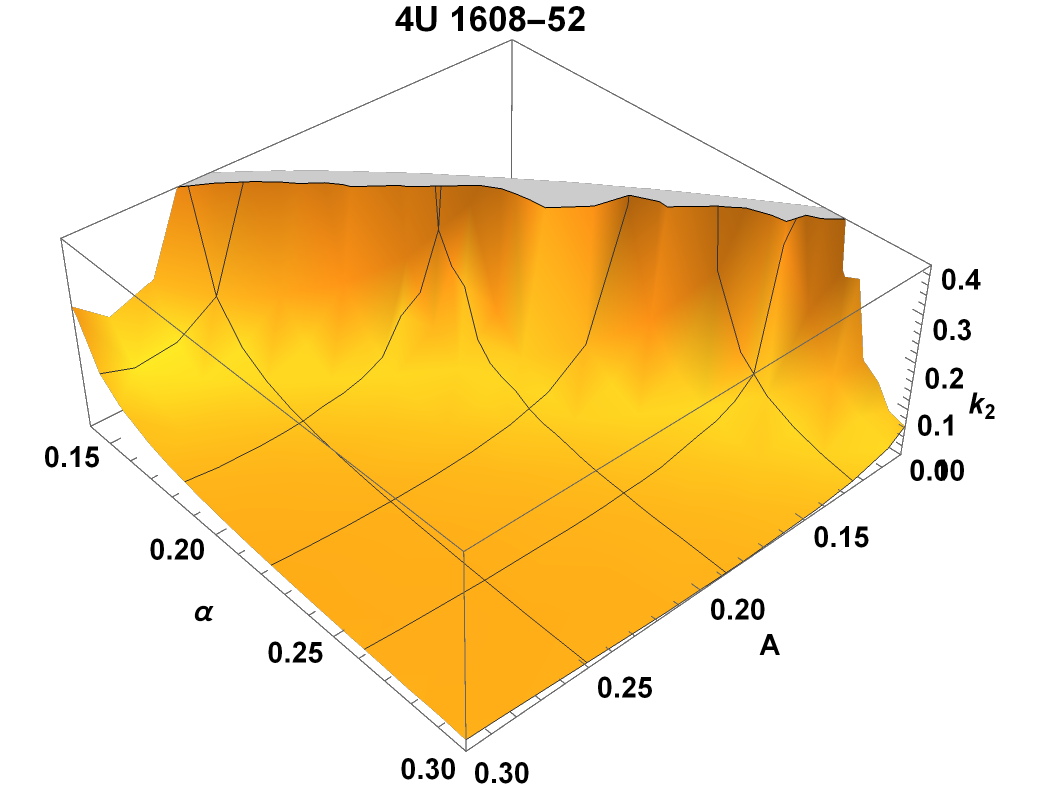}
\caption{Variation of Love number $k_2$ with respect to the parameter $\alpha$ and $A$ for the compact star $4U~1608-52$.}
 \label{fig:t13D}
  \end{figure}
 
\begin{figure}
\includegraphics[width=6.5cm]{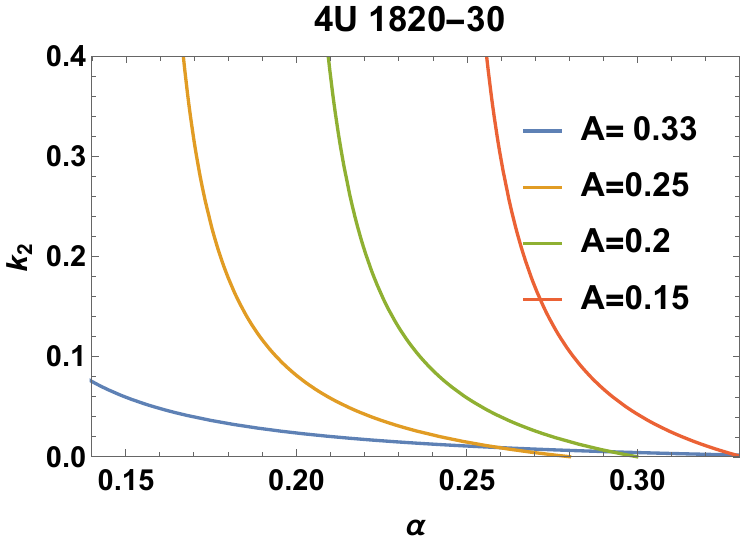}
\includegraphics[width=6.5cm]{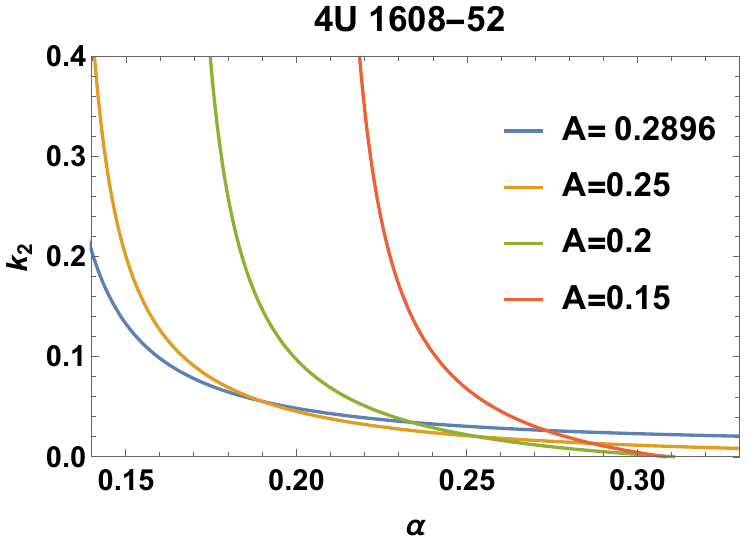}
\caption{Variation of Love number $k_2$ with respect to  $\alpha$ for the compact stars $4U~1820-30$ and $4U~1608-52$ under the specific choice of $A$.}
 \label{fig:t1}
  \end{figure}

\begin{figure}
\includegraphics[width=6.5cm]{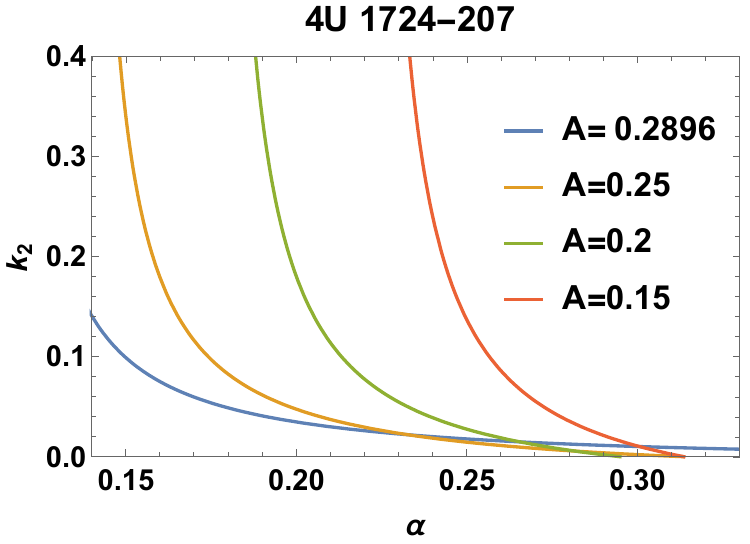}
\includegraphics[width=6.5cm]{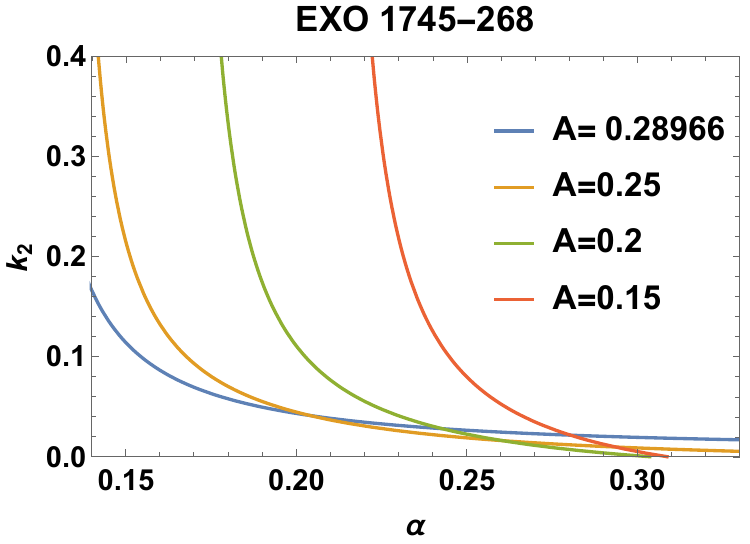}
\caption{Variation of Love number $k_2$ with respect to  $\alpha$ for the compact stars $4U~724-207$ and $EXO~1745-268$ under the specific choice of $A$.}
  	\label{fig:t2}
  \end{figure}

 \begin{figure}
\includegraphics[width=6.5cm]{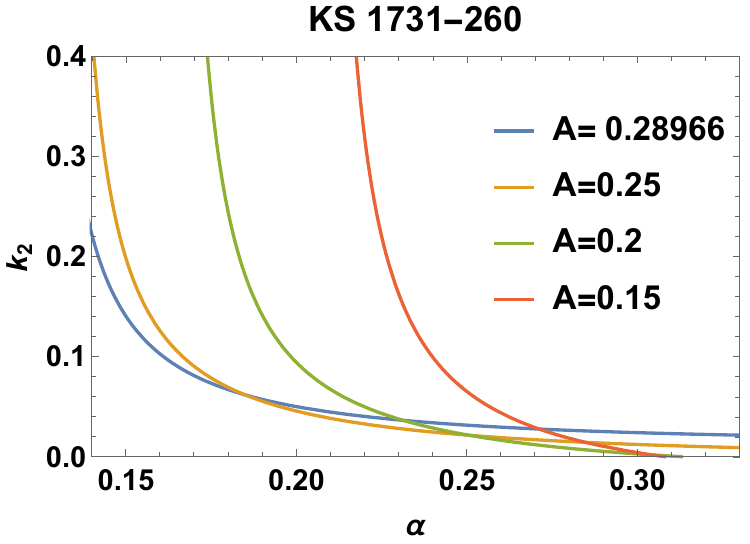}
\includegraphics[width=6.5cm]{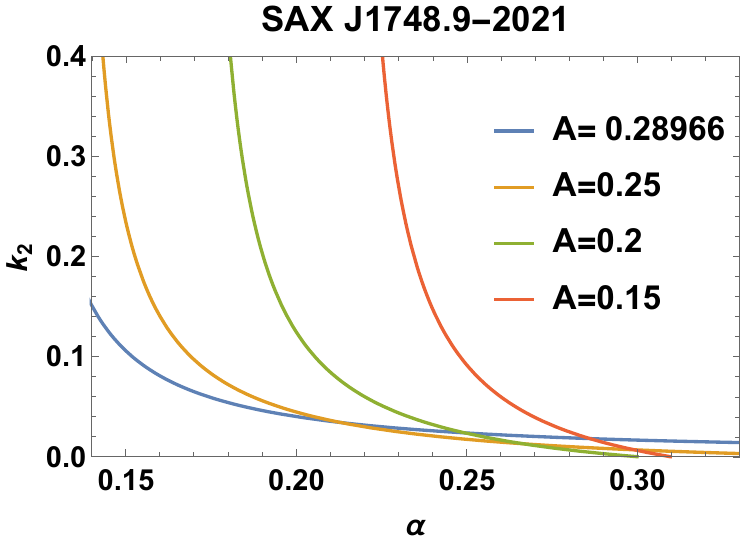}
\caption{Variation of Love number $k_2$ with respect to  $\alpha$ for the compact stars $KS~1731-260$ and $SAX~J1748.9-2021$ under the specific choice of $A$.}
  	\label{fig:t3}
  \end{figure}

In Table  \ref{tab1}, the numerical values of the tidal Love number $k_2$ are calculated and given for different compact stars with a random physically acceptable values of $\alpha$ and $A$.

\section{Discussion and Conclusion}\label{sec8}
In this paper, we have obtained a new class of interior solutions to the Einstein field equations for an anisotropic matter distribution obeying a linear EOS. We have done matching of the metrics along with the condition of the vanishing radial pressure at the boundary. Indeed, the smooth matching of two metrics on any hypersurface requires the continuity of both the first and the second fundamental form (Darmois conditions), from these conditions and the field equations it follows at once that the radial pressure (in absence of dissipation) must vanish on the boundary surface. In other words, vanishing of the radial pressure on the boundary is not and additional condition, but a consequence of the Darmois matching conditions. The solution has been shown to be regular as well as well-behaved and hence could describe a relativistic compact star. The anisotropy in the present model has been assumed to be due to tidal effect and hence we have calculated the Love number for several compact objects in 2$D$ as well as 3$D$ plots. 

In this work we have considered the metric potential $e^{\lambda}$ same as in the paper by Bhar et al.~\cite{Bhar2020}. However, in our paper we have used a linear EOS (Eq. 8) which is different from the EOS used in~\cite{Bhar2020}. On the other hand, in the paper~\cite{Das2020} the authors have used different metric potentials $e^{2\nu}$ and $e^{2\lambda}$ which is the anisotropic generalization of Korkina and Orlyanskii (solution III)~\cite{Korkina1991}, and hence it is different from our paper. In the present paper, as EOS is linear (different from the paper of~\cite{Bhar2020}), we get different expressions for the physical quantities $\rho,~p_r,~p_t$ and $\Delta$. Because of the differences in these physical quantities, the expressions of the quantities, viz. the mass function $m(r)$, energy conditions,  casuality condition, relativistic adiabetic index, TOV equation, Harrison-Zeldovich-Novikov's stability condition -- all are different from the works~\cite{Bhar2020,Das2020}. Similarly, though the expression of $e^{\lambda}$ resembles with~\cite{Maurya2019}, but that for $e^{\nu}$ is different. Hence, here also the expression of the physical quantities are different from our calculations.

Some of the salient features of the present model are as follows:

(i) To show that the solution can be used as a viable model for observed compact sources, we just consider as a specific example, the pulsar $4U~1608-52$ whose mass and radius are estimated to be $M = 1.57_{-0.29}^{+0.30}~M_\odot $ and $R = 9.8_{-1.8}^{+1.8}~$km, respectively~\cite{Ozel2016,Roupas2020}. For the given mass and radius, we have determined the values of the constants $a=0.011165$, $b=-0.0000191011$ and $A=0.289667$ for arbitrarily chosen values of $\alpha=0.22$, $\beta=-0.002$.  For physical acceptability of our model, using the values of the constants and plugging the values of $G$ and $c$, we have figured out the behaviour of the physically relevant quantities graphically within the stellar interior.

(ii) Fig.~\ref{fig:metrices} (left panel) shows that the metric potentials are positive within the stellar interior as per the requirement. On the other hand, Fig.~\ref{fig:metrices} (right panel) shows  the variations of the energy density $\rho$, radial pressure  $p_r$ and tangential pressure $ p_t $, respectively. The pressures are radially decreasing outwards from its maximum value at the centre and in case of radial pressure it drops to zero at the boundary as it should be, however the tangential pressure remains non-zero at the boundary. Obviously, all the quantities decrease monotonically from the centre to the boundary. In Fig.~\ref{fig:Density_figure} fall-off behaviour of the energy density and pressure have been shown satisfactorily, respectively in the left and right panels.

(iii) The mass function is shown graphically in Fig.~\ref{fig:Ani_figure} (left panel) from which one can note that the mass function is regular at the center. The variation of the anisotropy has been shown in Fig.~\ref{fig:Ani_figure} (right panel) which is zero at the centre as expected and is maximum at the surface. 

(iv) One important criterion of satisfaction of energy conditions is shown graphically in Fig.~\ref{fig:Strongenergycondition_figure} (left panel). The matching of the interior space-time with that of the exterior metric is shown in Fig.~\ref{fig:Strongenergycondition_figure} (right panel). 

(v) In the present work we have discussed the stability condition via the adiabatic index, however Heintzmann and Hillebrandt \cite{HH} showed that the presence of anisotropy in the system slows down the growth of instability and the increase of anisotropic factor changes the stability condition as $\Gamma > 4/3$. So, Moustakidis \cite{Moustakidisa,Moustakidisb} proposed a stricter condition known as critical value of the adiabatic index. This critical condition can be found whether our model is stable or not which obviously demands a detailed work and can be executed in a future project.

(vi) In the present model the origin of the anisotropy is considered from the gravitational tidal effects which caused deformation in the structure of the matter distribution. We therefore have calculated Love number for six different compact stars as shown in the Table \ref{tab1}. Here $k_2$ is approximately 0.03 to 0.1 for almost all the stars with the numerical values of parameters $\alpha$ = 0.22, $A=0.28$. It is interesting to note that our 2$D$ graphical plots resemble to those of the work by Yazadjiev et al.~\cite{Yazadjiev} in connection to the tidal Love numbers of neutron stars in $f(R)$ gravity. It is also to note that our computed values are nearer to the values reported elsewhere in different context which ranges from 0.111 to 0.207~\cite{Hinderer2008,Kramm2011}. 

It is worthy to note here that instead of comparing the numerical values of $k_2$, we have shown that the range of $k_2$ in our work is similar with~\cite{Bhar2020,Das2020}, because of the fact that we do not know the exact numerical values of the parameter $\alpha$ for different stars. Under this situation, in Table 1 we have chosen random numerical values of the parameters $\alpha$, $\beta$, $A$, $a$ and $b$ for which all other conditions are satisfied for different compact stars (see Figs. \ref{fig:metrices}--\ref{fig:dmdrho_figure}). The same technique has been performed in~\cite{Bhar2020,Das2020} though the parameters are different there from the parameters of our present paper.

Moreover, the procedure for the calculation of the tidal Love number is same for all these papers~\cite{Bhar2020,Das2020} as well as in the present paper. So, the derivation for the expression of the tidal Love number looks similar in these papers.  But one may notice that we have calculated tidal Love number from the modified master equation Eq. \eqref{t17}. The expression for $k_2$ contains the term $\mathit{y} = \left.\dfrac{r H'(r)}{H(r)}\right|_{r=R}$. The expression of $\mathit{y}$ has been calculated from Eq. \eqref{t20} which contains the term $e^{\lambda},~p_r,~p_t,~\rho$ etc. and hence different from other papers~\cite{Bhar2020,Das2020}. This aspect is evident from our Figs. \ref{fig:t13D}--\ref{fig:t3} all show the variation of $k_2$ w.r.t. $\alpha$ for different values of the parameter $A$. It is clear from these figures that with increasing $\alpha$ values of $k_2$ decreases. 

However, one interesting point has been highlighted by Kramm et al.~\cite{Kramm2011} that an observational $k_2$ would imply a maximum possible core mass and metallicity. In connection to the Neptune-sized exoplanet $GJ~436b$ especially for $k_2 < 0.24$, it would not help to further constrain interior models because in that regime the solutions are too degenerate. However, a $k_2 > 0.24$ would indicate a maximum core mass $M_c < 0.5~M_p$ and large outer envelope metallicities. 

As a final comment we would like to add here that the present study specially employs the following two concepts, viz. (1) we have presented a compact stellar model where the pressure anisotropy has been considered as the direct effect of the gravitational tidal deformation and (2) following Kramm et al.~\cite{Kramm2011} it has been highlighted that an observational $k_2$ would imply a maximum possible core mass and metallicity which obviously shed light regarding internal structure of the compact stars as well as invokes observational evidences in support of this idea. \\

\section*{Appendix A}

\begin{align}
\nu' =& \dfrac{1}{a r^2+b r^4+1}\left[r \left(a^2 \alpha r^2+a^2 r^2+2 a b \alpha r^4\right.\right.\nonumber\\
&+8 \pi  \beta \left(a r^2+b r^4+1\right)^2+2 a b r^4+3 a \alpha\nonumber\\
&\left.\left.+a+b^2 \alpha r^6+b^2 r^6+5 b \alpha r^2+b r^2\right)\right]. \label{soln}
\end{align}

Integrating we have
\begin{align}
\nu &=\frac{1}{12} r^2 \left[3 (\alpha+1) \left(2 a+b r^2\right)+8 \pi  \beta \left(3 a r^2+2 b r^4+6\right)\right]\nonumber\\
&+\alpha \log \left(a r^2+b r^4+1\right) +\text{c},\label{soln1}
\end{align}
and hence
\begin{align}
 e^{\nu(r)} &=A \left(a r^2+b r^4+1\right)^\alpha \nonumber\\
\times &\exp \left[\frac{1}{12} r^2 \left(3 (\alpha+1) \left(2 a+b r^2\right)+8 \pi  \beta \left(3 a r^2+2 b r^4+6\right)\right)\right],\label{soln3}
\end{align}
where $A=e^{c}$ is a constant of integration. 

Consequently, the physical quantities are obtained as
\begin{align}
\rho =& \frac{r^2 \left(\text{a}^2+5 \text{b}\right)+2 \text{a} \text{b} r^4+3 \text{a}+\text{b}^2 r^6}{8 \pi \left(1+\text{a} r^2+\text{b} r^4\right)^2},\label{den}\\
p_r =&\frac{\alpha \left(r^2 \left(\text{a}^2+5 \text{b}\right)+2 \text{a} \text{b} r^4+3 \text{a}+\text{b}^2 r^6\right)}{8 \pi \left(1+\text{a} r^2+\text{b} r^4\right)^2}+\beta, \label{radpres} \\
p_t =& 2 \pi  \beta^2 r^2 \left(a r^2+b r^4+1\right)+ \dfrac{1}{32 \pi} \left[a (\alpha+1)^2\right.\nonumber\\
&+b (\alpha+1)^2 r^2+\frac{4 (\alpha-3) \alpha r^2 \left(a^2-4 b\right)}{\left(a r^2+b r^4+1\right)^3}\nonumber\\
&+\frac{a \left(-4 \alpha^2+6 \alpha-2\right)+4 b (\alpha (2 \alpha-5)-1) r^2}{\left(a r^2+b r^4+1\right)^2}\nonumber\\
&\left.+\frac{(\alpha+1) \left(3 a \alpha+a+b (7 \alpha+3) r^2\right)}{a r^2+b r^4+1}\right]\nonumber\\
&+\dfrac{1}{2 (1+a r^2 + b r^4)}\left[\beta\left(a^2 (\alpha+1) r^4+2\right.\right.\nonumber\\
&\left.\left.a r^2 \left(2 b (\alpha+1) r^4+3 \alpha+4\right)+b (\alpha+1) r^4 \left(b r^4+5\right)\right)\right],\label{tangpres}\\	
\Delta &= \frac{1}{32}\left[16 \beta r^2 (a \alpha+a+4 \pi  \beta)+64 \alpha \beta+32 \beta\right.\nonumber\\
&+16 \beta r^4 (4 \pi  a \beta+b \alpha+b)+64 \pi  b \beta^2 r^6\nonumber\\
&+ \frac{4 (\alpha-3) \alpha r^2 \left(a^2-4 b\right)}{\pi  \left(a r^2+b r^4+1\right)^3}+\frac{(\alpha+1)^2 \left(a+b r^2\right)}{\pi }\nonumber\\
&+\frac{4 b (\alpha (2 \alpha-9)-1) r^2-2 a \left(2 \alpha^2+\alpha+1\right)}{\pi  \left(a r^2+b r^4+1\right)^2}\nonumber\\
&+\dfrac{1}{\pi( a r^2+b r^4+1)} \left[-32 (2 \pi  \alpha+\pi ) \beta\right.\nonumber\\
&\left.\left.a \left(3 \alpha^2-16 \pi  (2 \alpha+1) \beta r^2+1\right)+b (\alpha (7 \alpha+6)+3) r^2\right] \right]. \label{ani}
\end{align}

\section*{Appendix B}

\begin{align}
F_g &= \frac{r}{16 \pi  \left(a r^2+b r^4+1\right)^3} \left[ a^2 r^2 \left(\alpha+8 \pi  \beta r^2+1\right)\right.\nonumber\\
&+8 \pi  \beta \left(b r^4+1\right)^2+b r^2 \left(b (\alpha+1) r^4+5 \alpha+1\right)\nonumber\\
&\left.+a \left(2 b (\alpha+1) r^4+16 \pi  b \beta r^6+3 \alpha+16 \pi  \beta r^2+1\right)\right] \nonumber\\
&\times \left[a^2 r^2 \left(\alpha+8 \pi  \beta r^2+1\right)+8 \pi  \beta \left(b r^4+1\right)^2\right.\nonumber\\
&\left.+b (\alpha+1) r^2 \left(b r^4+5\right)+a(3 (\alpha+1)\right.\nonumber\\
&\left.\left.2 b (\alpha+1) r^4+16 \pi  b \beta r^6++16 \pi  \beta r^2\right)\right],
\end{align}

\begin{align}
F_a &=  \dfrac{r}{16} \left[16 \beta (a \alpha+a+4 \pi  \beta)+\frac{b (\alpha+1)^2}{\pi }\right.\nonumber\\
    &+16 \beta r^2 (4 \pi  a \beta+b \alpha+b)+64 \pi  b \beta^2 r^4+ \dfrac{4 \chi_{1}}{\pi \eta^3}\nonumber\\
    &\left. + \dfrac{2 \chi_{2}}{\pi \eta^2}+ \dfrac{\chi_{3}}{\pi \eta}  \right],\\
  F_h &= \dfrac{1}{4 \pi \left(a r^2+b r^4+1\right)^3} \alpha r \left[5 a^2-5 b+b^3 r^8\right.\nonumber\\
    &\left.+a r^2 \left(a^2+13 b\right)+3 b r^4 \left(a^2+4 b\right)+3 a b^2 r^6\right],   
\end{align}
where $\eta = a r^2+b r^4+1$, $\chi_{1} = (\alpha-3) \alpha \left(a^2-4 b\right)$, $\chi_{2}= a^2 \left(2 \alpha^2+\alpha+1\right)+a b \left(2 \alpha^2+\alpha+1\right) r^2+2 b (\alpha (2 \alpha-9)-1)$, $\chi_{3} = a^2 (\alpha+1)^2+a b (\alpha+1)^2 r^2+16 \pi  a (2 \alpha+1) \beta+b \left(7 \alpha^2+32 \pi  (2 \alpha+1) \beta r^2+6 \alpha+3\right)$.\\

\section*{Appendix C}

To determine $k_2$, consider background metric $^{(0)}g_{\mu \nu}(x^{\nu})$ of a compact object. With linear perturbation  $h_{\mu \nu}(x^{\nu})$ of the background metric, the modified perturbed metric can be written as 
 	\begin{eqnarray}
 		g_{\mu \nu}\left(x^{\nu}\right)=^{(0)} g_{\mu \nu}\left(x^{\nu}\right)+h_{\mu \nu}\left(x^{\nu}\right). \label{t3}
 	\end{eqnarray}
 	
We write the background geometry of the spherical static star in the standard form
 	\begin{eqnarray}
 		^{\left(0\right)} ds^{2} =^{\left(0\right)} g_{\mu \nu }dx^{\mu }dx^{\nu } =-e^{\nu(r) }dt^{2}+e^{\lambda(r) }dr^{2}+r^{2}\left( d\theta ^{2}+\sin ^{2}\theta d\phi ^{2}\right). \label{t4}
 	\end{eqnarray}
 	
With this linearized metric perturbation $h_{\mu \nu}(x^{\nu})$, following the works in the refs.~\cite{Regge1957,Biswas2019}, we restrict ourselves to static $l = 2, \, m=0$ even parity perturbation. With these restriction the perturbed metric becomes
\begin{eqnarray}
 		h_{\mu \nu}=\operatorname{diag}\left[H_{0}(r) e^{\nu}, H_{2}(r) e^{\lambda}, r^{2} K(r), r^{2} \sin ^{2} \theta K(r)\right] Y_{2 0}(\theta, \phi), \label{t5}
 	\end{eqnarray}
 where $H_0,~H_2$ and $K$ are radial functions determined by perturbed Einstein equations.
 	
 	For the spherically static metric with anisotropic pressure, the stress-energy tensor is given as~\cite{Bowers1974,Doneva2012,Herrera2013,Pretel2020}
 	\begin{eqnarray}
 		^{(0)}T_{\chi}^{\xi}=\left(\rho+p_{t}\right) u^{\xi} u_{\chi}+p_{t} g_{\chi}^{\xi} - \left(p_{t}-p_{r}\right) \eta^{\xi} \eta_{\chi}, \label{t6}
 	\end{eqnarray}
 	where $u^{\chi} u_{\chi} = -1$, $\eta^{\chi}\eta_{\chi}=1$ and $\eta^{\chi} u_{\chi} = 0$. 
 	
 	Furthermore, the perturbed stress-energy tensor is defined as $T_{\chi}^{\xi} = ^{(0)}T_{\chi}^{\xi} + \delta T_{\chi}^{\xi}$. The non-zero components of perturbed stress-energy tensor are: $\delta T_{t}^{t} = - \dfrac{d\rho}{dp_{r}} \delta p_{r} Y(\theta ,\phi ), \\ \delta T_{r}^{r} = \delta p_{r}(r) Y(\theta ,\phi ),$ \& $  \delta T_{\theta}^{\theta} = \delta 	T_{\phi}^{\phi} =  \frac{dp_{t}}{dp_{r}}  \delta p_{r}(r) Y(\theta ,\phi ).$	
 	
 	With these perturbed quantities we can write down the perturbed Einstein field equation which becomes (where as already mentioned that $G=c=1$)
 	\begin{eqnarray}
 		G_{\chi}^{\xi} = 8 \pi T_{\chi}^{\xi}. \label{t7}
 	\end{eqnarray}
 	
 The non-zero components of the  background Einstein field equation gives the expressions
 \begin{align}
 	^{(0)}G_{t}^{t} &= 8 \pi ^{(0)}T_{t}^{t}\nonumber\\
 	\Rightarrow \lambda'(r) &= \frac{8 \pi  r^2 e^{\lambda (r)} \rho (r)-e^{\lambda (r)}+1}{r}, \label{t8}\\
 	^{(0)}G_{r}^{r} &= 8 \pi ^{(0)}T_{r}^{r}\nonumber\\
 	\Rightarrow \nu'(r) &= \frac{8 \pi  r^2 p(r) e^{\lambda (r)}+e^{\lambda (r)}-1}{r}. \label{t9}
 \end{align}
 
 Also, we know that $\nabla_{\xi}^{(0)}T_{\chi}^{\xi} = 0 $. Therefore, choosing $\xi = r$, by expanding and solving the equation, we can find the expression as 
 \begin{eqnarray}
 	p'(r) = \frac{1}{2 r}\left[-4 p(r)+4 p_{t}(r)-r p(r) \nu^{\prime}(r)-r \rho(r) \nu^{\prime}(r)\right]. \label{t10}
 \end{eqnarray}
 
 The various components of perturbed part of the Einstein field equation \eqref{t7} gives these expressions
 \begin{eqnarray}
 	&G_{\theta}^{\theta} - G_{\phi}^{\phi} = 0 \Rightarrow H_{0}(r) = H_{2}(r) = H(r), \label{t11}\\
 	&G_{r}^{\theta} =0 \Rightarrow K' = H' + H \nu', \label{t12}\\
 	&G_{\theta}^{\theta} + G_{\phi}^{\phi} = 8 \pi (T_{\theta}^{\theta} + T_{\phi}^{\phi}) \Rightarrow \delta p = \frac{H(r) e^{-\lambda (r)} \left(\lambda '(r)+\nu '(r)\right)}{16 \pi r \frac{dp_{t}}{dp}}. \label{t13}
 \end{eqnarray}
 	
 Using the identity, $\dfrac{\partial^{2} Y(\theta,\phi)}{\partial \theta^{2}} + cot(\theta) \dfrac{\partial Y(\theta,\phi)}{\partial \theta}+ \csc ^2(\theta ) \dfrac{\partial^{2} Y(\theta,\phi)}{\partial \phi^{2}} = -6 Y(\theta ,\phi )$ as well as Eqs. \eqref{t8} - \eqref{t13}, we have the master equation for $H(r)$ as
 	\begin{eqnarray}
 		&- \frac{1}{e^{-\lambda (r)} Y(\theta ,\phi )} \left[ G_{t}^{t} - G_{r}^{r}\right] = - \frac{8 \pi}{e^{-\lambda (r)} Y(\theta ,\phi )} \left[ T_{t}^{t} - T_{r}^{r}\right]\nonumber\\
 		&\Rightarrow  H''(r) + \mathcal{R} H'(r) + \mathcal{S} H(r) = 0, \label{t14}
 	\end{eqnarray}
where
 	\begin{eqnarray}
 		\mathcal{R} = - \left[ \frac{-e^{\lambda (r)}-1}{r}-4 \pi  r e^{\lambda (r)} (p_{r}-\rho (r))\right], \label{t15}
 	\end{eqnarray}
 	
 	\begin{eqnarray}
 	\mathcal{S} =- \left[16 \pi  e^{\lambda (r)} \left(p_{r} \left(e^{\lambda (r)}-2\right)-p_{t}(r)-\rho (r)\right)\right. \nonumber\\
 		+ \left. 64 \pi ^2 r^2 p_{r}^2 e^{2 \lambda (r)}+\frac{4 e^{\lambda (r)}+e^{2 \lambda (r)}+1}{r^2}\right. \nonumber\\
 		+ \left.\frac{-4 \pi  \frac{d\rho}{dp_{r}} e^{\lambda (r)} (p_{r}+\rho (r))-4 \pi  e^{\lambda (r)} (p_{r}+\rho (r))}{\frac{dp_{t}}{dp_{r}}}\right]. \label{t16} 
 	\end{eqnarray}
 	
 	The expressions for $e^{\lambda},~\rho(r),~p_{r}(r),$ and $p_{t}(r)$ can be found from the solutions of a physically acceptable model.
 	
 	The tidal Love number $k_2$ can be calculated by matching the internal solution with the external solution of the perturbed variable $H(r)$ at the surface of the star~\cite{Bhar2020,Das2020}. Then the modified expression of tidal Love number can be found in terms of $y$ and compactness $\mathcal{C} = M/R$ as 	
 	\begin{eqnarray}
 		k_2 = [8 (1-2 \mathcal{C})^2 \mathcal{C}^5 (2 \mathcal{C} (\mathit{y}-1)-\mathit{y}+2)]/X, \label{t17}
 	\end{eqnarray}
 where
 	\begin{align}
X &= 5(2 \mathcal{C} (\mathcal{C} (2 \mathcal{C} (\mathcal{C} (2 \mathcal{C} (\mathit{y}+1)+3 \mathit{y}-2)-11 \mathit{y}+13)+3 (5 \mathit{y}-8))\nonumber\\
&\left.-3 \mathit{y}+6) 
+3 (1-2 \mathcal{C})^2 (2 \mathcal{C} (\mathit{y}-1)-\mathit{y}+2) \log \left(\frac{1}{\mathcal{C}}-2\right)\right.\nonumber\\
&\left.-3 (1-2 \mathcal{C})^2 (2 \mathcal{C} (\mathit{y}-1)-\mathit{y}+2) \log \left(\frac{1}{\mathcal{C}}\right)\right).  \label{t18}
 	\end{align}
 
Here $\mathcal{C} = \frac{M}{R}$ and $ \mathit{y}$ depends on $r,\, H$ and its derivatives
 	 \begin{eqnarray}
 	 	\left. y = \dfrac{r H'(r)}{H(r)}\right|_{r=R}.  \label{t19}
 	 \end{eqnarray}
 	 
In order to get the numerical of $k_2$ for a particular compactness $\mathcal{C}$, let us modify Eq. \eqref{t14}, by using Eq. \eqref{t19}, as~\cite{Rahmansyah2020a}
 	 \begin{eqnarray}
 	 	r \mathit{y}' + \mathit{y}^2 + (r \mathcal{R} -1) \mathit{y} + r^2 \mathcal{S} = 0. \label{t20}
 	 \end{eqnarray}

{\bf CRediT authorship contribution statement }\\
Shyam Das: Conceptualization. B.K. Parida: writing-original draft. Saibal Ray: writing- 
review and editing. Maxim Khlopov and K.K. Nandi: supervision. \\

{\bf Declaration of competing interest }\\
The authors declare that they have no known competing financial interests or personal relation- 
ships that could have appeared to influence the work reported in this paper.\\

\section*{Acknowledgement}
SD and SR gratefully acknowledge support from the Inter-University Centre for Astronomy and Astrophysics (IUCAA), Pune, India under its Visiting Research Associateship Programme. Research of MK was supported by the Ministry of Science and Higher Education of the Russian Federation under Project ``Fundamental problems of cosmic rays and dark matter", No. 0723-2020-0040.

\end{document}